# Thermal Energy Grid Storage Using Multijunction Photovoltaics

Caleb Amy[1], Hamid Reza Seyf[2], Myles A. Steiner[3], Daniel J. Friedman[3], Asegun Henry[1,2]

[1]Department of Mechanical Engineering, Massachusetts Institute of Technology,
Cambridge MA, 02139, USA

[2]George W. Woodruff School of Mechanical Engineering, Georgia Institute of Technology,
Atlanta GA, 30332, USA

[3]National Renewable Energy Laboratory,
Golden CO, 80401, USA

## Abstract

As the cost of renewable energy falls below fossil fuels, the most important challenge to enable widespread sustainable power generation has become making renewables dispatchable. Low cost energy storage can provide this dispatchability, but there is no clear technology that can meet the need. Pumped hydroelectric and compressed air storage have low costs, but they are geographically constrained. Similarly, lithium-ion batteries are becoming ubiquitous, but even their lower bounding asymptote cost is too high to enable cost-competitive dispatchable renewables. Here, we introduce a concept based on thermal energy grid storage (TEGS) using a multijunction photovoltaic heat engine (MPV) with promising initial experimental results that could meet the low cost required to enable cost competitive dispatchable renewables. The approach exploits an important tradeoff between the accession of an extremely low cost per unit energy stored, by storing heat instead of electricity directly, while paying the penalty of a lower round trip efficiency. To understand why this tradeoff is advantageous, we first introduce a framework for evaluating storage technologies that treats round trip efficiency (RTE) as a variable, in addition to cost per unit energy stored (CPE) and cost per unit power (CPP). It is from this perspective that the TEGS-MPV concept offers a compelling economic proposition.

## Introduction

In the last decade the cost of electricity derived from renewables i.e., solar photovoltaics (PV) and wind, has fallen dramatically[1,2] making renewables cheaper or cost competitive with fossil derived electricity in ideal locations. This is a remarkable achievement, but it is based purely on an assessment of the levelized cost per unit energy (LCOE) (i.e., the total cost divided by the lifetime electrical energy output, $/kWh-e). Although this is an important quantity, it does not account for the fact that renewable electricity is not necessarily provided when desired, since it is inherently tied to the weather. Thus, dispatchability remains a key necessity provided by existing fossil-based technologies. Consequently, as Denholm[3,4] and others[5-7] have shown, renewable penetration onto the grid will be limited to < 10-15% without grid level storage. Thus, "the storage problem" i.e., how to store/buffer energy at the grid scale cheaply, has emerged as one of the most important technological barriers to decarbonization of the grid and mitigating climate change.

Currently the cheapest grid storage technology is pumped hydroelectric (PH), which has a high roundtrip efficiency (RTE) ~80-90%, as well as a low cost per unit energy (CPE) ~$75/kWh-e and low cost per unit power (CPP) ~$1/W-e[8]. The issue with PH, and also compressed air energy storage (CAES), however, is that it is geographically limited and, in the case of PH, the prime locations have already been exploited[7,9-11]. Electrochemical batteries, on the other hand, have promising new chemistries[7,12], but it is unclear if any will displace Li-ion batteries whose prices continue to drop from $300-400/kWh-e down to a predicted asymptote ~ $150/kWh-e[7]. There is significant concern nonetheless, that even this lower asymptote for Li-ion is still not cheap enough to enable the eventual 100% penetration of renewables needed to truly mitigate climate change. In this respect, alternative solutions to the storage problem are needed, and it is likely that costs closer to $50/kWh-e and below[13,14] will be needed to eventually realize 100% penetration and full abatement of $CO_2$ emissions from the stationary power sector. This low cost requirement arises from the fact that a levelized cost of storage (LCOS) below the $0.06/kWh average electricity price[15] and 10 or more hours[16] of storage are needed to reliably and cost-effectively supply the grid.

In thinking about lower cost storage, one class of technologies that has not received much attention is thermal energy storage (TES). This is because the final form of energy needed is electricity, necessitating the conversion of heat back to electricity, which tends to occur at low efficiency (~35-40%) for conventional turbine-based heat engines. It is because of this low efficiency, that the idea of taking electricity off the grid, converting it to heat e.g., via joule heating, storing it and then converting it back to electricity has seemed nonsensical. However, even though the low efficiency is off-putting, when one considers the entire economic proposition, it can actually still prove quite attractive – especially when new embodiments that achieve somewhat higher RTEs, or very low CPEs[17] and/or CPPs are considered.

Several embodiments[18,19] have been proposed and are under development involving the conversion of electricity to heat, which is then stored and later converted back on demand, such that we have generally termed this class of technologies thermal energy grid storage (TEGS) herein. What these various incarnations share is the storage of heat, which is exploited to be as inexpensive as possible, and it should be noted here that storing heat can be an order of magnitude cheaper than electrochemical batteries. The simplest embodiment that is arguably closest to commercialization, is to use molten salt as is currently done in concentrated solar power (CSP) plants[20], except that one would need to replace the solar heat input with joule heating. With this approach, one can today achieve a CPE < $100/kWh-e[21], but the problem would be the low RTE (~35-40%). A more clever approach introduced by Laughlin[18] involves the usage of a heat pump instead of joule heating, which can in theory almost double the RTE to ~ 72%, and makes TEGS a very attractive option. Other interesting and potentially attractive embodiments also exist, but to determine the best option, the value of RTE must be assessed with respect to CPE and CPP. It is therefore important to have a framework for quantitatively evaluating the tradeoffs between RTE, CPE and CPP, which ultimately dictate the economics and value to the grid. In what follows, we briefly introduce a simple framework for assessing such tradeoffs, followed by an introduction and discussion of our own incarnation of TEGS, which our analysis shows may be one of the few

solutions to the storage problem that is inexpensive enough to eventually enable a fully renewable grid.

For a given storage technology, the total capital expenditure (CAPEX) can be thought of as a sum of two main components, $CAPEX = CPE + CPP/t$ where $t$ is the time that the resource can be discharged at maximum power. In the simplest terms, neglecting operating expenditures, one must compare the CAPEX to the two primary unsaturated sources of revenue, namely capacity payments ($/kW) paid annually, which scale with the power output promised/supplied, and arbitrage ($/kW) earned annually, which is where the RTE plays a critical role. Sioshansi et al.[4] have quantified how much value a storage resource would receive from arbitrage, as a function of the RTE and $t$, by using the Pennsylvania New Jersey Maryland (PJM) grid as an example. Their work showed that there would be diminishing value for large $t$ resources on the 2007 PJM grid and they quantified how the value of storage changes with RTE, which we have used as an input in Fig. 1A. This plot shows that a storage technology with RTE ≤ 36% would not have generated any value from arbitrage on the 2007 PJM grid. Fundamentally, this is because energy must be purchased and therefore the ratio between on-peak and off-peak pricing sets a lower efficiency limit to earn arbitrage profit, as shown in Eq. 1. That is, if a technology must buy three times as much energy as it sells, it must sell that energy for at least three times the purchase price to derive positive value from arbitrage. Because these devices cannot charge or discharge instantaneously, the closer their efficiency is to $\eta_{min}$, the less frequently they can profitably engage in arbitrage.

$$\eta_{min} \approx \frac{P_{offpeak}}{P_{peak}} \quad (1)$$

$$CPP = \frac{L}{\alpha}(V_{arb}(RTE) + CP) - t \times CPE \quad (2)$$

$$\alpha = \frac{(r \times t)e^{(r \times t)}}{e^{(r \times t)} - 1} \quad (3)$$

To assess the value of RTE relative to CPE and CPP, we can use the simple relation in Eq. 2 to assess the tradeoffs between RTE, CPE and CPP. Here, the CPP for zero net present value (NPV) is evaluated where the total expenditure is equivalent to the total revenue earned during the system's life, discounted with an internal rate of return (IRR) of 10%, denoted $r$. Here, $L$ is life in years, $V_{arb}(RTE)$ is arbitrage value in $ kW$^{-1}$ yr$^{-1}$, which is a function of the RTE, as shown in Fig. 1A. The capacity payment ($CP$) is estimated based on the average net cost of new entry (Net CONE) of peaking gas turbines. Net CONE is the cost of a peaking gas turbines minus its anticipated revenue from energy sales and ancillary services, and therefore it represents the capacity payment it must be paid to break even[22,23]. For the results in Fig. 1, $CP$ is taken to be $95/kW and the details associated with how this was calculated are given in the Methods section. Future revenue is discounted with the factor $\alpha$ as shown in Eq. 3, which assumes revenue is accrued uniformly over the life of the system.

Using this relationship, the maximum value of CPP that can be tolerated for a given technology under this simplified scenario was calculated, assuming the values given in the table of Fig. 1B. Justification for these values are detailed in the Methods section. Using this framework, any storage technology can be evaluated by knowing its RTE, CPE and CPP. By using its actual CPE value (horizontal axis) and corresponding RTE (color) from Fig. 1C, one can read off the maximum allowable CPP for the technology on the vertical axis. If it turns out that a given technology's actual CPP value is lower than the corresponding max CPP value in Fig. 1C, then it would be profitable under the stated financial assumptions. The max CPP and actual CPP values for different technologies are then indicated in Fig. 1C, as well as the estimated values for the technology introduced herein, which could be profitable.

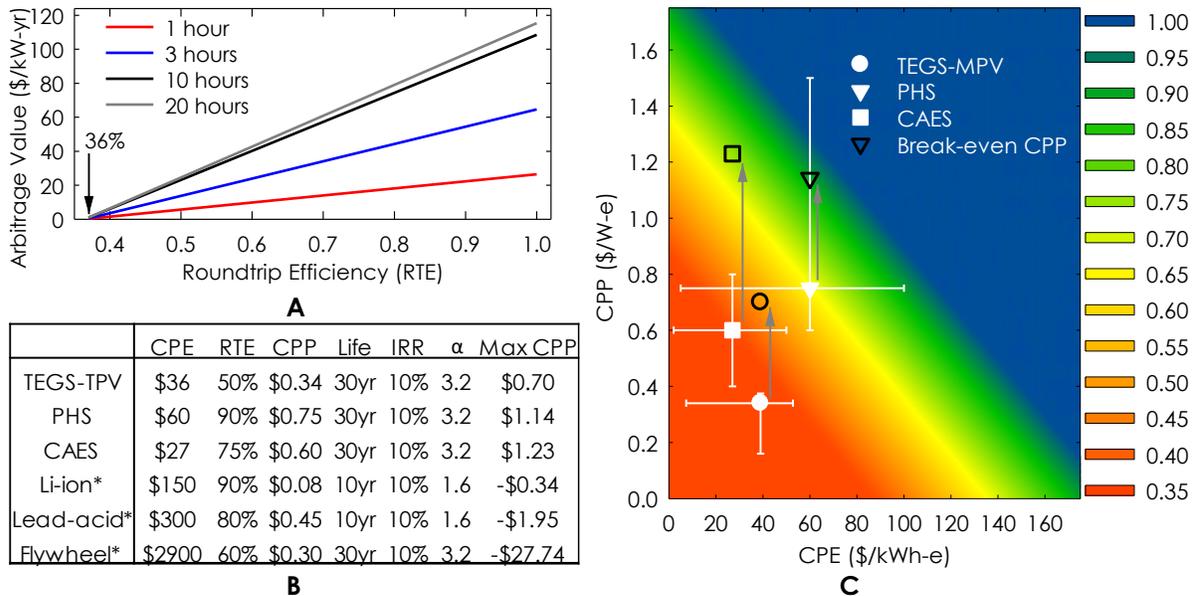

**Fig. 1:** Three parameter analysis of energy storage value: CPE, CPP, and RTE. **A** – Value of Arbitrage as a function of RTE. **B** – Value comparison of leading energy storage technologies (* indicates not shown in C, because they are off the chart). **C** – CPE and CPP (white shapes) of three competitive energy storage technologies. Arrows and black shapes indicate maximum CPP to break even. Arrow direction and length indicates NPV. The colored contour represents the RTE to break even, assuming 10% IRR, 30 year system, and 10 hours of storage.

The results in Fig. 1C show that although it is initially non-intuitive to operate at low RTE, economically it makes sense to still consider low RTE technologies that have very low CPE and CPP values. For example, a storage technology with an RTE of 50%, CPE < $50/kWh-e and CPP < $0.5/W-e can be profitable like previous installations of PH, while batteries with 10 year life, RTE of 90%, CPE of $150/kWh-e, and CPP of $0.08/W-e would not be profitable (i.e., under the stated assumptions in Fig. 1B). Thus, based on the economic motivation in Fig. 1, we introduce a new TEGS concept termed TEGS-MPV that employs TES at ultra-high temperatures to achieve an RTE $\geq$ 50% and a CPE < $50/kWh-e, and most notably uses multijunction photovoltaics (MPV) to achieve a CPP < $0.5/W-e. With this approach, a storage technology that is not geographically limited, yet has similar cost effectiveness to PH could be realized and could become the most cost-effective embodiment of TEGS.

## A New System Concept

The new TEGS-MPV system concept is illustrated in Fig. 2 and consists of a low cost thermal storage fluid, nominally 553 metallurgical grade (98.5% pure) silicon, which costs ~ $1.6/kg at high volume. The liquid Si is stored in a "cold" tank, nominally at 1900°C, in the discharged state. To charge the system, the 1900°C Si is pumped, using an all graphite seal-less sump pump, through a series of pipes that are externally irradiated by graphite heaters that draw electricity from the grid. In this heater sub-system, the temperature of the Si is nominally raised to ~2400°C as it is pumped into the "hot" tank, where it is stored. In this process, excess electricity from the grid is stored as sensible heat in the liquid over a 500°C temperature difference (1900-2400°C). The tanks are large, with diameters on the order of 10 m, which allows the surface area to volume ratio to be small enough that less than 1% of the energy stored is lost each day, which is similar to CSP plants using molten salt TES[24]. In this extreme temperature case, the insulation is more expensive as detailed in the Cost Modeling section, but heat loss can still be minimal. Assuming such a storage resource were to be discharged once a day, this leads to a small and almost negligible penalty on the RTE. When electricity is desired, the 2400°C Si is pumped out of the hot tank and through a MPV power cycle. The MPV power cycle is envisioned to consist of an array of graphite pipes that are covered in tungsten (W) foil. The W foil acts as a lower vapor pressure barrier between the graphite pipes and the MPV cells, which are mounted to an actively cooled block that keeps their temperature near the ambient temperature (i.e., ~30°C). The W foil therefore serves as a photon emitter, almost identical to an incandescent lightbulb[25] that emits light to the MPV cells, which subsequently convert a fraction of it to electricity. As the Si passes through the graphite piping network it cools down to ~1900°C, as energy is extracted and converted to electricity, at which point it is returned to the "cold" tank to await later recharging. Here, it is important to note that for this temperature range, 25-33% of the light being converted is in the visible spectrum, but materials with band gaps more optimal than silicon solar cells are envisioned. Therefore, these cells are arguably just PV cells, as opposed to thermophotovoltaic (TPV) cells. For this reason, we've elected to use the term multijunction PV (MPV) to highlight the fact that the envisioned cells bear resemblance to, and use many of the advances that have been made for MPV in the context of concentrated PV (CPV).

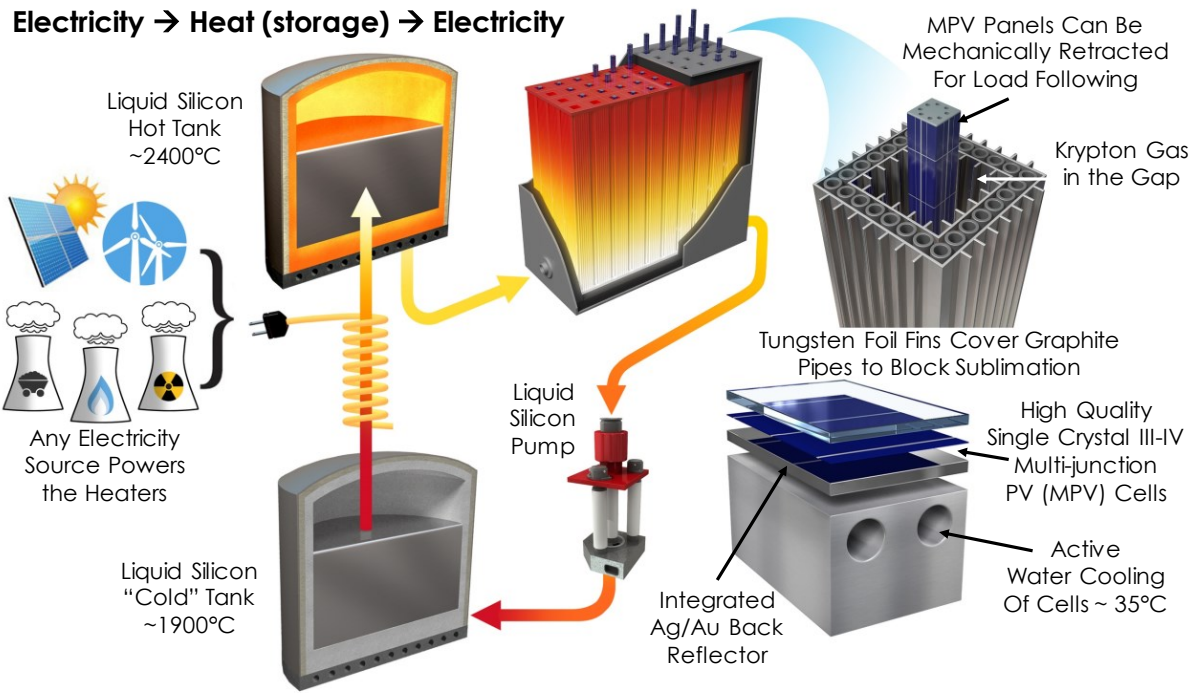

**Fig. 2:** Schematic overview of the proposed technology, where electricity from any source is converted to heat via joule heating, which is then transferred to a liquid storage medium as sensible heat (1900-2400°C). Using a cheap material, e.g., metallurgical grade silicon ($1.6/kg), the heat can then be stored cheaply, with minimal heat leakage (~ 1%/day) at large scale. When electricity is desired, the liquid is pumped through an array of tubes which emit light. The light/heat is then converted back into electricity using an array of multi-junction photovoltaic cells that convert the visible and near infrared light, but use an integrated mirror on the back surface to reflect back the unusable light.

There are several novel and important features associated with this embodiment of TEGS that will first be expounded upon, before the next section where feasibility is discussed. Most notably, it should be appreciated the temperature regime chosen here is essentially the practical upper limit for industrially manufactured refractories, namely graphite and W. Although both materials melt/decompose at even higher temperatures, 2400°C represents a rather practical limit due to the substantial vapor pressure that develops, which can lead to emitter material deposition onto the MPV cells, which would degrade their optical performance. Nonetheless, the reason for employing the most extreme temperatures possible is to achieve the highest possible RTE.

As detailed in the next section, to the best of our knowledge, the efficiency of PV that converts light from a high temperature heat source, most commonly referred to as TPV, has not exceeded ~29%, which was achieved using single junction silicon cells and a 2000°C emitter[26]. However, more recent work using an InGaAs cell achieved almost the same efficiency (28.8%)[27], but only required an emitter temperature of 1250°C. This is because the single most important loss in such cells is the voltage, which is usually ~ 0.3-0.4 V lower than the band gap voltage for III-V materials[28]. However, this loss is worse for a material such as silicon, because of its indirect band gap as well as Auger recombination at the high photon fluxes in our application. Given that this voltage loss tends to be almost constant when practical devices are made (neglecting a small bandgap dependence), it suggests that the most important pathway to reaching higher efficiencies (> 40%) with a terrestrial light source, is to utilize higher band gap materials than have been pursued previously. In this way, the rather fixed voltage loss, becomes a smaller fraction of the band gap voltage, and the overall device efficiency improves at higher band gaps. However, higher band gaps require higher temperatures to ensure that a substantial portion of the emitted spectrum is above the band gap and can be converted. This is important, because to achieving a low CPP requires that the cells be operated at high power density, so that their cost, which scales with the total cell area and is inversely proportional to the power output per unit area, can remain low. Furthermore, although photons below the band gap can be returned to the emitter, this recycling is imperfect, so the proportion of photons above the band gap must be substantial to outweigh the below band gap absorption. It is for these reasons that the most extreme temperatures are considered in this TEGS embodiment.

Although it is appreciated that such a concept may seem unrealistic, our recent work[29] and the initial experiments above 2000°C described herein have laid the foundation for the high temperature infrastructure in Fig. 2, by addressing some of the most critical risks. Key to this approach is the use of a flowing liquid as a heat storage and heat transfer medium. This is because it enables constant (steady state) power output and efficiency as opposed to inherently transient output associated with a medium such as a solid or phase change material. Usage of a solid or formation of a solid (i.e., via PCM) can become problematic because a sensible heat discharge from a solid will inherently cause a conductive resistance to build up, which will lower the power output over time, and will cause the heat engine efficiency/RTE to decay with time as well. This issue, which is expounded upon in the Methods, is undesirable for grid operation, because as previously noted, capacity payments are based on a rated power that can be supplied/promised. However, since a liquid/fluid can be pumped, it enables straightforward designs that can achieve a steady state power output and RTE at a peak power output.

For the MPV power cycle, some of the important system level considerations have been addressed in previous work by Seyf and Henry[30], such as the need for the power cycle to be large (MW scale, with length scales ~ 10 m) in order to overcome the losses associated with heat leakage to the environment by minimizing the ratio of surface area to volume. Their prior work also identified the back surface reflector (BSR) reflectivity or more specifically the net amount of below band gap cell absorption, as the most critical parameter. However, while their initial predictions for the cell efficiency are theoretically justified, they do not accurately capture the realistic voltage losses that tend to occur in real cells. Thus, a more realistic consideration of practical cell losses would drive the system towards operation at much higher temperatures than their initial work indicated[30], as will be discussed in more detail in the next section. Nonetheless, the MPV cells considered herein are still envisioned to incorporate a BSR, but for this temperature regime, higher band gap materials as well as multiple junctions are expected to be optimal. Furthermore, by using cells grown on GaAs as opposed to InP substrates, and using hydride vapor phase epitaxy (HVPE) instead of metal-organic (MOCVD) chemical vapor deposition, cell costs could be much lower than what is estimated herein.

One critical question that arises with the TEGS-MPV approach, however, is why MPV is chosen as the heat engine instead of a turbine, which could likely be more efficient at lower temperatures. There are three reasons for this: (1) Turbines that take an external heat input and operate in this temperature range (i.e. > 1000°C) do not currently exist. Although it may be possible to develop such a system, for turbines there is a large barrier to commercial deployment, as it would require a large OEM to undertake an expensive > $100M development effort for a high-risk application. On the other hand, existing III-V cell manufacturers are positioned to facilitate the commercialization and deployment of the described MPV power cycle with much less investment. (2) The cost of our proposed MPV system can be much lower than that of a turbine. (3) The speed with which turbine-based heat engines can ramp from zero to full power is on the order of tens of minutes to an hour. However, with this TEGS-MPV approach, as is illustrated in Fig. 2, the MPV modules can be actuated in and out of the light on the order of seconds, which could provide much greater value to the grid, via load following, thereby increasing revenue.

## Modeling and Experimental Results on Feasibility

One inescapable component needed to realize the TEGS-MPV system is the storage medium tank. If there is no conceivable way to make the storage tanks, then there is no path forward towards realizing the liquid-based TEGS-MPV system. Using a liquid storage medium requires that the tank be impermeable, and the options for materials at these temperatures are severely limited. One of the only cost effective options is graphite, but it would be infeasible to fabricate the entire 10-30 m diameter tank from a single monolithic piece. This necessitates that the tank be formed from sections, with sealed interfaces that do not leak.

A first and highly encouraging result that suggests this problem can be easily and cost effectively solved, is shown in Fig. 3. In this experiment, a dense (1.85 g/cc) graphite (KYM-20) miniature "tank" filled with 553 grade Si was heated above 2,000°C for 60 minutes. The tank was made from two sections and sealed with a thin grafoil face seal that was compressed by carbon fiber composite (CFC) threaded rod and nuts. The tank was insulated with graphite felt and aluminum silicate insulation inside a quartz tube, under high purity Argon gas (< 1 x $10^{-6}$ atm $O_2$). The tank was heated by induction and its temperature was measured using a C-type thermocouple. It is well known that graphite and Si(l) react to form SiC[31], and a protective SiC scale that prevents further reaction can form, if the graphite has the right microstructure. In this experiment, as shown in Fig. 3E, Si penetrated the graphite tank approximately 400 μm, and created a dense 20 μm thick SiC layer at the interface, preventing further penetration. Initial experiments showed that if a thicker grafoil seal is used, the expansive reaction would break apart the entire "tank". Thus, this preliminary result is rather non-trivial, as it offers initial proof and confidence that Si can be contained at these temperatures in a multi-section tank.

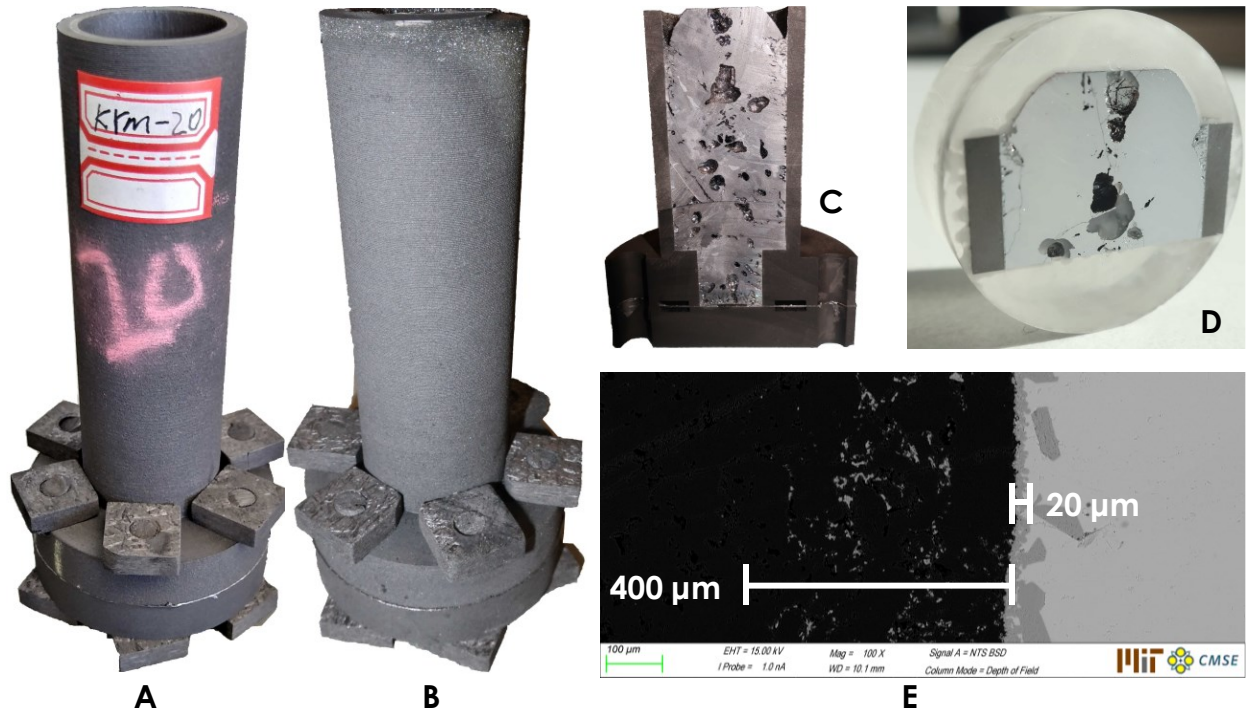

**Fig. 3:** Proof of concept experiment demonstrating a sealed graphite (KYM-20) reservoir containing 553 grade Si above 2,000°C for 60 minutes. Shown before (**A**), after (**B**), cross sectioned (**C**), and polished (**D**). **E** – SEM backscattered image of tank wall, showing SiC protective layer.

Another feasibility issue concerns the need to pump liquid Si at temperatures as high as 2400°C. On this issue, recent experiments by Amy et al.[29] have shown that it is feasible to use brittle ceramics as mechanical pumps, as they pumped liquid tin at temperatures up to 1400°C. Here, the temperature is ~1000°C higher, which would be a major concern for an infrastructure made from solid metal components. However, for ceramics and refractories, such as graphite, this is much less of a concern, since the materials tend to be covalently bonded and therefore tend to exhibit weak dependence of their mechanical properties on temperature. In fact, the strength of graphite actually increases with temperature, up to 2600°C[32]. For these reasons, although pumping at 2400°C has not been done before, the testing above 2000°C and pumping[29] at 1400°C renders the notion now feasible, as there are no obvious issues that should prevent operation at the higher temperatures.

Another potential issue with the TEGS-MPV system is that the heater efficiency is a non-trivial matter, since it would require power conditioning electronics that could have substantial losses that ultimately detract from the RTE. This issue is discussed in more detail in the methods section, but the conclusion of the analysis is that existing power electronics are capable of supplying the necessary inputs at low cost with less than 1% parasitic loss[33]. This loss is small and almost negligible, which then shifts our focus to the primary loss in the system, which occurs in the MPV power block.

The PV cell which converts the photons radiated from the thermal emitter to electrical energy plays a crucial role in the system efficiency and almost entirely dictates the RTE. The energy in a photon ($eV_{photon}$) incident on the cell can suffer several types of losses, which are very strongly dependent on the spectrum of light incident on the cell, and on the design of the cell itself. The most significant is the voltage loss, where an incident photon is absorbed by the cell and converted into electrical current, at a cell open-circuit voltage $V_{OC} < V_{photon}$. This energy loss $E_{loss} = eV_{photon} - eV_{OC}$ can be partitioned into two individual losses related to the junction bandgap $E_g$: $E_{loss} = (eV_{photon} - E_g) + (E_g - eV_{OC})$. The first loss, referred to as thermalization loss, arises because the thermally-radiated spectrum contains a wide range of photon energies, so that no one junction bandgap is precisely at the photon energy for all photons in the spectrum. To mitigate this loss, we envision using a two-junction photovoltaic device, with the two bandgaps chosen to optimally convert a band of the spectrum. This MPV approach, illustrated schematically in Fig. 4A, is well established for solar CPV, and is the only effective approach that has been demonstrated to mitigate thermalization losses.

The second aspect of the voltage loss, which is typically ~ 0.3–0.4 eV for high-quality (Ga,In)(As,P)-based III-V devices at conventional operating conditions of ~25°C and one-sun photon flux, is also unavoidable. This loss is largely due to an increase in entropy when photons enter the cell from a small solid angle and are reradiated into a much larger solid angle[34]. An additional voltage penalty occurs, due to non-radiative recombination, and this penalty is greater for cells made from silicon than from III-V materials, due to silicon's indirect bandgap. It should be noted nonetheless that silicon was used in the experiments that led to the highest reported TPV efficiency we're aware of (~29%[26]), and therefore by using III-V materials, it is an important step toward mitigating this penalty and reaching higher efficiencies. Most importantly, however, is the fact that the $E_g - eV_{OC}$ penalty is proportionally smaller for higher bandgap materials than for lower bandgap materials. The very high emitter temperatures used in the TEGS-MPV concept generate correspondingly high-energy photons for which relatively high-bandgap PV cells are suitable, thus mitigating the $E_g - eV_{OC}$ penalty.

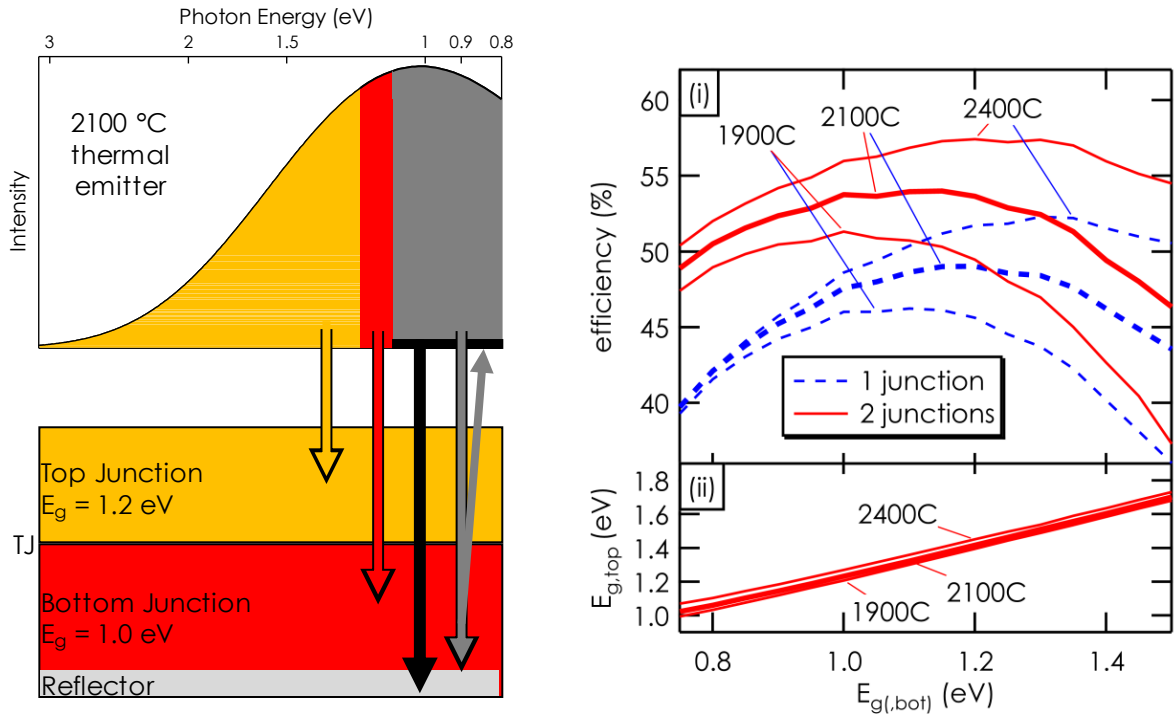

**Fig. 4A** – Schematic illustration of the absorption of various bands of the spectrum from a 2100°C thermal emitter in a two-junction PV cell. The reflectivity of the cell for photon energies below the 1-eV bottom-junction bandgap is assumed to be 98%, meaning that 98% of the sub-bandgap photons (gray color in the figure) are returned to the source, while 2% (black in the figure) are absorbed in the back reflector. "TJ" indicates the tunnel junction series interconnect. **B** – (i) Modeled efficiencies of 1- and 2-junction PV cells for 1900°C–2400°C emitter temperatures as a function of (bottom) junction bandgap for the 1-junction cells. For 2-junction cells, the top-junction bandgap is selected to give the highest efficiency for the given bottom-junction bandgap. (ii) Optimal top-junction bandgap for the 2-junction cells.

A type of absorption loss also occurs at the system level and results from the free carrier absorption of sub-bandgap photons, either at the BSR or in the semiconductor itself. The BSR should be designed to be as reflective as possible, so that the unusable sub-bandgap photons are reflected back to the thermal emitter to keep it hot, which is a critically enabling feature for this system concept[30]. A reflectivity of at least 98%, as illustrated in gray in Fig. 4A, is necessary to realize the high performance predicted in Fig. 4B. The additional losses associated with series resistance and shadowing are discussed in the Methods section.

Fig. 4B shows the modeled efficiencies of 1-junction and series-connected 2-junction cells as a function of junction bandgap, for a range of thermal emitter temperatures. The cell modeling was performed using the very well experimentally validated model of Geisz *et. al.*[35] including a conservative but realistic $E_g/e$-$V_{OC}$ = 0.4 V penalty. The incident spectrum was computed using tungsten's emissivity[36] and the diffuse gray band approximation to account for the MPV cells high absorptivity above the band gap and low absorptivity below the band gap. Consistent with the cell measurements, we assumed 2% of the light below the cell's band gap is absorbed, which is illustrated in the 2100°C spectrum is shown in Fig. 4A. Every above-bandgap photon which is absorbed is assumed to be collected as current, an idealization that can later be replaced with actual measured cell performance. Nonetheless, this assumption is close to the measured performance of many previous cells. The cell's current-voltage characteristic and maximum-power output are then computed, with junction voltages adding for multijunction devices. The ratio of this power output to the integrated net input power, including a static 4.6 kW/m$^2$ convection loss (see Methods section for details) through the inert gas between the emitter and cell, yields the net cell efficiency. Practical cell efficiencies are typically ~85-90% of the efficiencies modeled at this level of idealization[35]. Fig. 4B shows that practical efficiencies of well over 40% are achievable for 1-junction cells, and ≥ 50% is possible with 2-junction cells. For the 2100°C emitter, the optimal junction bandgaps are roughly 1.0-1.2 eV for the 1-junction cell, and {1.2, 1.0} eV to {1.4, 1.2} eV for the {top, bottom} junctions of a 2-junction cell. Using a dual junction PV device, fabricated using the inverted metamorphic multijunction (IMM) cell architecture (additional details in the Methods section), these calculations show that ≥ 50% RTE is possible with TEGS-MPV.

## Cost Estimation

The major advantages of TEGS-MPV over other grid level energy storage technologies, such as PH, are its estimated low cost (less than half PH) and the fact that it is not geographically limited — which is the primary drawback of PH and CAES. Thus, it is important to demonstrate how we arrived at the cost estimates provided. Fig. 5 shows bar chart break-downs of the various costs. As a nominal design point, we considered a TEGS-MPV system rated at 100 MW-e peak output with 10 hours of storage. The CPE includes the cost of the storage medium, tank, insulation, auxiliary components, and construction, using a similar procedure to Glatzmaier[17] as well as Wilk et. al.[37], which is described in the Methods section. The CPP includes the cost of the heater, MPV cells, inverter, emitter, insulation, construction, and cooling system. Following a summary of the basis of these costs, the Methods section describes the methodology used to generate these estimates and associated sources.

In the base case, 553 grade (98.5% pure) Si is used at a market price of $1.60/kg. The tank wall is made from isostatic molded graphite (e.g. KYM-20) of density 1.8g/cm$^3$, at a cost of $7/kg based on multiple quotes from large suppliers. The insulation for all components consists of graphite felt ($7,000/m$^3$), surrounded by aluminum silicate blanket, surrounded by fiberglass blanket. The cost of the graphite felt dominates, so its use is constrained to the region above the 1,350°C temperature limit of aluminum silicate. Construction costs are based on the cost of molten-salt CSP plants[37], plus the labor cost of assembling additional components as detailed in the methods. The cost of the heater includes graphite heating elements, graphite pipes and headers, insulation, and inert containment. The MPV power block contains similar elements, although the cost is dominated by the $0.08/W-e inverter cost[1], $0.10/W-e MPV cell cost, and $0.07/W-e cooling cost. This cell cost is based on an assumed power density of 100 kW/m$^2$ and a cell cost[30] of $10,000/m$^2$. In reality, this cost may be much lower if the aforementioned cell manufacturing developments are realized (i.e., GaAs substrates and HVPE). In this less conservative lower cost case, the MPV cell cost is negligible and the inverter dominates. Similarly, an alternative embodiment of interest is if Fe partially, or fully replaces Si as the storage medium. In this alternative scenario, the cost of the medium becomes extremely low if one uses scrap steel, and the other tank costs, especially insulation and construction, dominate. In this less conservative

lower cost case, we also assume a lower grade extruded ($2/kg) graphite is used for the tank and a higher heat loss of 2% per day, instead of 1%. These changes to the design affect the system cost as shown in Fig. 5.

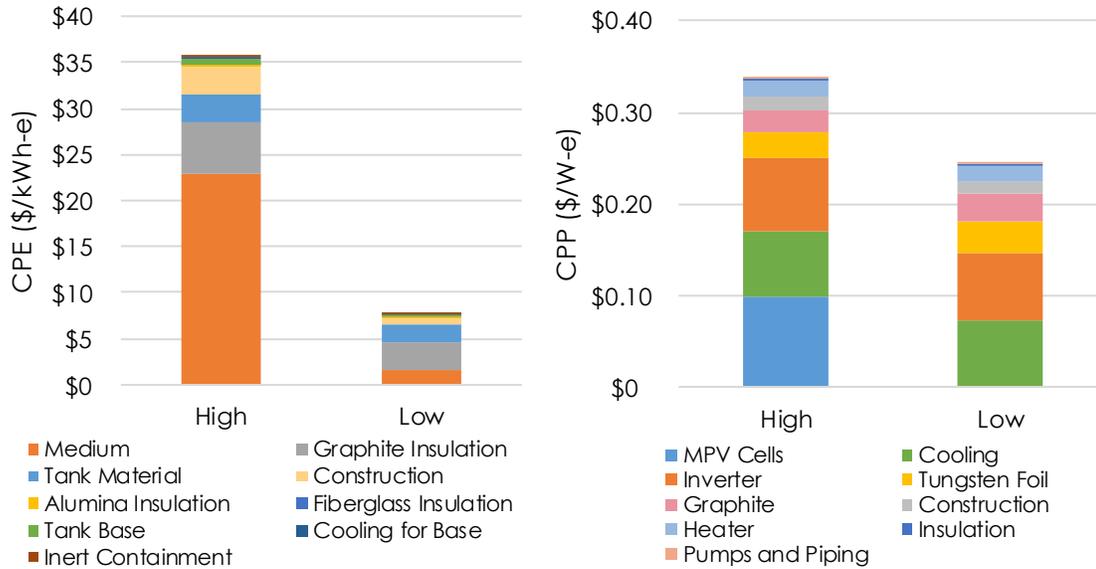

**Fig. 5A** – Estimated CPE of TEGS-MPV in base case and low cost case. **B** – Estimated CPP of TEGS-MPV in base case and low cost case. Low cost case assumes scrap steel as storage medium, and PV cell cost near silicon cells.

## Conclusions

Based on the analysis presented herein, the TEGS-MPV concept offers an attractive value proposition as a grid storage technology. In this study, a simplified framework was presented that enables one to compare and assess the economic viability of new grid storage solutions. Of particular importance is the tradeoff space between RTE, CPE and CPP. The analysis showed that based on the previous work of Denholm[4], there is likely a RTE lower bound of ~ 36%, below which a grid storage technology may be unlikely to generate any revenue from arbitrage. Nonetheless, there are TEGS embodiments that have the potential to significantly exceed this RTE lower bound, while also achieving extremely low CPE and CPP. A new TEGS embodiment was then presented that involves the use of ultra-high temperature storage media, in liquid form, and uses MPV as the converter. This new approach has several noteworthy benefits including the ability to reach $\geq$ 50% RTE with a CPP < $0.5 W-e, and the potential to offer load following capabilities to grid operators. These benefits strongly suggest that if realized the TEGS-MPV approach could be one of the few grid storage approaches that are inexpensive enough to enable the eventual 100% penetration of renewables onto the grid. However, there are a number of remaining practical challenges that must be overcome to realize the potential cost and performance presented herein. Most notably a prototype system is needed to confirm that the storage medium can be reliably pumped, no leaks form anywhere in the flow loop, and the MPV cells can be reliably fabricated and tested under the described conditions. Towards this end, however, two first experiments were presented that strongly suggest some of the most risky aspects of the TEGS-MPV system can be resolved. First, experiments above 2000°C showed that a tank for Si with

dimensions on the order of 10 m, could conceivably be made out of smaller (i.e., order 1 m) sections that are sealed and bolted together. These experiments showed that grafoil gaskets can be used to successfully seal against liquid silicon without leakage. Additionally, the most important property of the MPV cells is their absorptivity for below band gap radiation. Calculations herein assumed this parasitic absorption to be 2% and measurements of the reflectivity of cells that were backed by a gold or silver layer have confirmed that this is indeed possible[27]. For these reasons, it seems feasible, although very challenging, that one could potentially realize the cost and performance described herein (i.e., CPE < $50/kWh-e, CPP < $0.5W-e, and RTE $\geq$ 50%) using TEGS-MPV.

**Methods:**

## Value Modeling:

The value of an energy storage system is expected to increase greatly in the future, especially as low-cost variable renewable energy begins to become available in excess of demand. However, in order to evaluate the merits of energy storage technologies, the value is estimated based on current market conditions, where the two revenue streams unlikely to be saturated are capacity payments and arbitrage[3]. Here, arbitrage value is defined as the summation of the annual revenue that could be earned by a device, minus the cost to purchase energy at off-peak times[4].

The revenue from capacity payments (CP) is highly variable, mainly because these payments are used to incentivize new generation capacity, and are therefore low in markets with excess capacity, and high in markets short on capacity[23]. Thus, instead of taking an average of actual CPs, which have spreads from $0-500/kW-yr[3], a more fundamentally robust method is used to estimate the CP that one could expect to earn. That is, the reason a CP is offered in the first place is that in regulated markets, grid resources that only operate during peak times (namely peaking gas turbines) do not earn enough revenue from energy sales to be profitable. Then, logically, a CP can be expected to be the subsidy needed to allow a balancing resource to have zero NPV. This is exactly what the net cost of new entry (Net CONE)[38] parameter represents, and it is widely reported. This cost is calculated as the total cost of a resource, minus the revenue it earns from energy sales and other ancillary revenues. Thus, it is this net loss that needs to be compensated by a CP. Here, this value is estimated to be $95/kW-yr based on the 2018 average Net CONE in the PJM market[22].

## Multijunction Photovoltaics (MPV)

### Additional Losses - Resistive and shadowing loss:

Practical cells are subject to a voltage loss due to series resistance. At the several hundred kW/m$^2$ power densities of light incident on the PV cells envisioned herein, the cell's output current density would be well above 10 A/cm$^2$, high enough to require effective mitigation strategies for the series resistance losses. Such current densities are comparable to those encountered in high-concentration solar photovoltaics, and the same mitigation strategies are applicable, centering on the use of properly engineered front-contact grids to reduce the main source of series resistance in III-V PV cells. In general, raising the grid coverage lowers the series resistance loss at the expense of raising the shadowing loss, and an optimal tradeoff in grid coverage must be found to balance these two competing losses. For this MPV application, this tradeoff may be much less demanding than for concentrated solar PV (CPV), because many of the photons blocked from entering the cell by the front grids will be reflected back to the thermal emitter so that their energy can be reused. Finally, at the module level, series resistance losses may be mitigated (again using similar techniques as applied in CPV) by using a larger number of small cells connected in series for high-voltage low-current operation, rather than fewer larger cells.

## Manufacturing Approach

As described in detail in Geisz, *et. al.*[35], the top junction is grown first, followed by the bottom junction, which leaves the back contact layer accessible for fabrication of a high reflectivity BSR. The device is then bonded to a handle and the substrate removed. As shown in Extended Data Figure 1, 1.0-eV $Ga_{0.7}In_{0.3}As$ is lattice-mismatched with respect to the GaAs substrate, with ~2.1% larger lattice constant, and has already been demonstrated to be a successful high quality third junction in high efficiency four-junction solar cells [39], with a voltage penalty Woc< 0.4 V. Illustrated in Extended Data Figure 1B, the mismatched cell is fabricated by first growing a compositionally step-graded buffer layer (CGB) that incrementally increases the lattice constant and relieves the accumulating strain[40], leading to a mismatched cell with a low threading dislocation density. Among the design requirements of the CGB is the necessity for it to be transparent to light below the top junction bandgap, and while this is true for the CGBs in the IMMs of France and Geisz, *et. al.*[39,40], there is concern about free carrier absorption in those layers for this application. Therefore, instead, we envision growing a tandem cell by adding a 1.2-eV $Al_{0.15}Ga_{0.55}In_{0.3}As$ above the GaInAs cell, at the same lattice constant, and then completely removing the CGB during processing. The resulting two-junction device would be thin, mounted to a stable handle and have a high reflectivity BSR. Variations on this design are also possible to achieve other band gap combinations as well, and it is through this new design of a MPV that the TEGS-MPV concept can achieve $\geq$ 50% RTE.

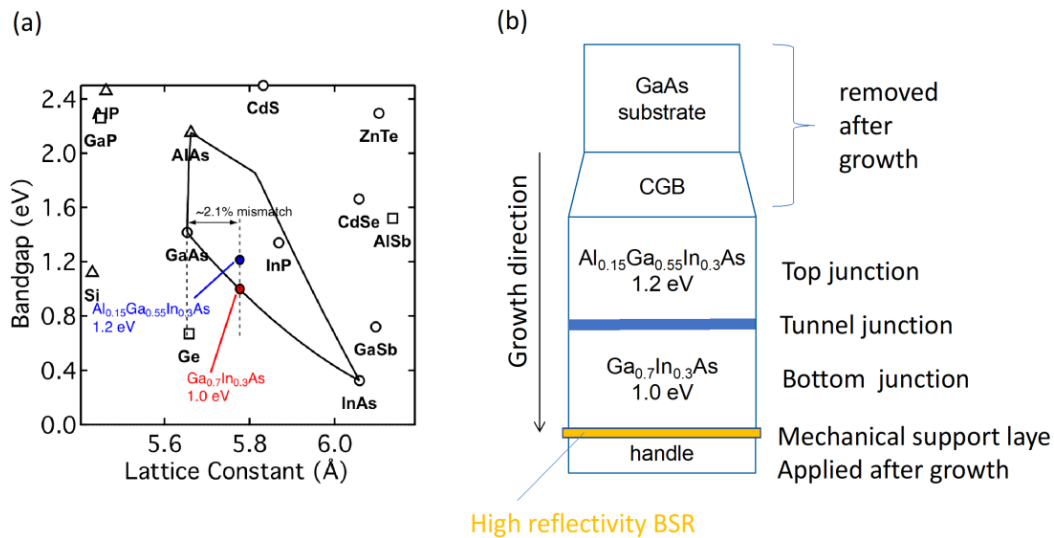

**Extended Data Figure 1A:** (a) Bandgaps and lattice constants of the common III-V binary and ternary compounds. The 1.2 and 1.0 eV alloys are indicated. (b) Schematic of the proposed cell design. The substrate and CGB layers would be removed during processing, leaving a thin two-junction device with a high reflectivity BSR.

## Radiative Heat Transfer in Optical Cavity

In this section, we quantify how the emitter to MPV cell surface area ratio ($A_{emitter}/A_{cell}$) changes the incident flux of light on the MPV cell. This is of interest because tungsten (W), which is needed to prevent evaporation of the emitter material, has a lower emissivity than graphite. In a design where the emitter and MPV have equal amount of area, this would yield a significant penalty on the output power density, which would translate to corresponding penalties on the efficiency and cost. Therefore, we quantify here how the total flux of photons with an energy greater than the band gap is affected by increasing the ratio of emitter to MPV cell surface area. As this ratio $A_{emitter}/A_{cell}$ is increased, the total radiative resistance between MPV cells and the emitter decreases. We used Monte Carlo Ray Tracing (MCRT) to study radiative heat transfer in the optical cavity and the reader is referred to Howell and Siegel[41] and Haji-Sheikh and Howell[42] for associated calculation details. To increase the $A_{emitter}/A_{cell}$, we examine a system where W foil fins are attached to W foil plates that cover the graphite pipes and therefore serve as the emitter. For simplicity, the calculation assumes the cell has back surface reflectivity of 0.98 for photons with energy below the bandgap. The reflectivity of MPV cell for photons with energy above the bandgap is assumed to be zero. The W emitter is assumed to be at 2100°C while the cell temperature is assumed to be 25°C. The spectral reflectivity of W is taken from the experimental data of Coblentz[36]. The distance between adjacent fins (D) and the total number of fins on each planer face of the emitter are assumed to be 2 cm and 8, respectively. These two parameters are kept fixed in the simulation while the length of fins ($L_{fin}$) and the size of MPV cells ($L_{cell}$) are changed from 8cm to 14 cm and 0 to 2.8 cm, respectively to achieve the required $A_{emitter}/A_{cell}$.

In the MCRT model, the number of computational cells on the surface of the optical cavity is controlled to contain roughly the same number of particles and consequently similar statistical variation. It was found that around 50 particles per computational cell on average is enough to resolve radiation field on the MPV cells and W emitter. Furthermore, the effect of the number of particles in the optical cavity was studied and it was found that depending on the size of optical cavity roughly 200,000 – 450,000 particles are sufficient to capture the radiation distribution on the surfaces of optical cavity. Therefore, 200,000 – 450,000 photon bundles were traced in each MCRT iteration, since tracing more photon bundles per time step did not change the simulation results, but significantly increased the computational time.

The effect of $A_{emitter}/A_{cell}$ on incident radiative power above the bandgap for two different bandgaps (Eg), i.e., 1.43 eV and 1eV is shown in Extended Data Figure 2. As expected, increasing $A_{emitter}/A_{cell}$ enhances the above bandgap incident power onto the cells. For Eg =1.43 eV, compared to the base case ($A_{emitter}=A_{cell}$), the incident power increases by a factor of 1.6 when the $A_{emitter}/A_{cell}$ =4. The analysis suggests, increasing the emitter to cell area is an effective approach to enhance the power output and consequently the efficiency.

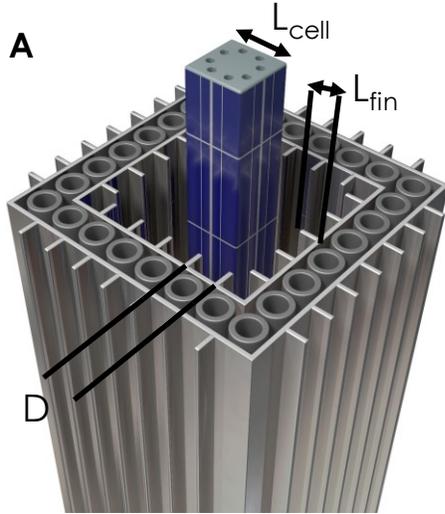
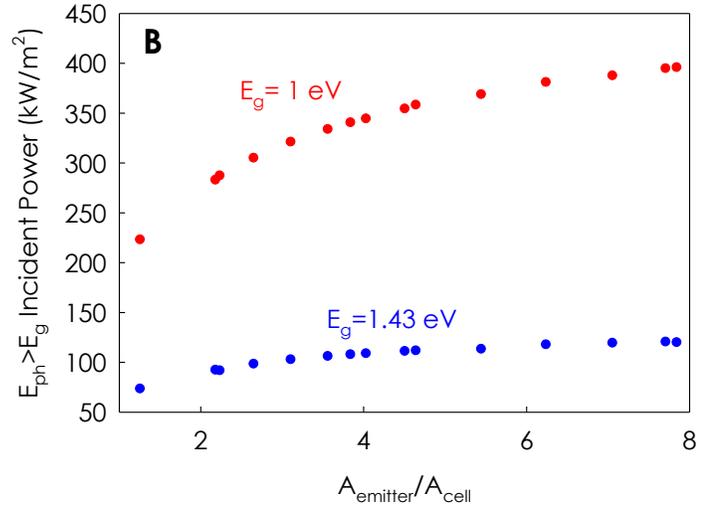

**Extended Data Figure 2: A –** Illustration of an individual power block sub-unit identifying the characteristic lengths used in the MCRT calculations. **B–** The effect of A$_{emitter}$/A$_{cell}$ on above bandgap incident radiative power.

## Convective Heat loss

Here, we analyze the heat loss from a single MPV sub-unit of the power block consisting of MPV cells surrounded by the planar W emitter with fins as an important first step towards determining the efficiency of the power block. The motivation of this analysis is twofold. First, although one can intuitively reason that usage of inert gas between the emitter and cell decreases the emitter material deposition onto the MPV cells, it is not clear *a priori* if the convective heat loss from the emitter is sufficiently low to enable to high power output. Second, it is expected that W fins on the emitter increases the effective emissivity of tungsten above the band gap of semiconductor and consequently the incident radiative power onto the MPV cells, but it is not clear if these fins enhance the convective heat loss from the emitter negating the gains. If the existence of W fins causes significant convective heat loss, then increasing the radiative power due to enhancement in emitter surface area may not be as effective overall. In this section, we describe a 3D steady-state flow and heat transfer model for evaluating the convective heat loss from the emitter so that this can be quantified.

The computational domain consists of planer W surfaces emitter with W fins and MPV cells. The total vertical height of the system and the distance between the fins are 2 m and 2 cm, respectively. The emitter temperature is assumed to be 2100°C while the MPV cell temperature is 25°C. The inert gas between the emitter and MPV is assumed to be krypton with temperature dependent viscosity, thermal conductivity and density. The flow is taken to be Newtonian, viscous and compressible. The governing equations are as follows

Continuity equation:

$$\frac{\partial(\rho u)}{\partial x}+\frac{\partial(\rho v)}{\partial y}+\frac{\partial(\rho w)}{\partial z}=0 \qquad (1)$$

Momentum Equations:

$$\frac{\partial(\rho u^2)}{\partial x}+\frac{\partial(\rho uv)}{\partial y}+\frac{\partial(\rho uw)}{\partial z}=-\frac{\partial P}{\partial x}+\mu\left(\frac{\partial^2 u}{\partial^2 x}+\frac{\partial^2 u}{\partial^2 y}+\frac{\partial^2 u}{\partial^2 z}\right) \quad (2)$$

$$\frac{\partial(\rho vu)}{\partial x}+\frac{\partial(\rho v^2)}{\partial y}+\frac{\partial(\rho vw)}{\partial z}=-\frac{\partial P}{\partial y}+\mu\left(\frac{\partial^2 v}{\partial^2 x}+\frac{\partial^2 v}{\partial^2 y}+\frac{\partial^2 v}{\partial^2 z}\right)+g(\rho-\rho_a) \quad (3)$$

$$\frac{\partial(\rho wu)}{\partial x}+\frac{\partial(\rho wv)}{\partial y}+\frac{\partial(\rho w^2)}{\partial z}=-\frac{\partial P}{\partial z}+\mu\left(\frac{\partial^2 w}{\partial^2 x}+\frac{\partial^2 w}{\partial^2 y}+\frac{\partial^2 w}{\partial^2 z}\right) \quad (4)$$

Energy Equation:

$$\frac{\partial(\rho C_p uT)}{\partial x}+\frac{\partial(\rho C_p vT)}{\partial y}+\frac{\partial(\rho C_p wT)}{\partial z}=k\left(\frac{\partial^2 T}{\partial^2 x}+\frac{\partial^2 T}{\partial^2 y}+\frac{\partial^2 T}{\partial^2 z}\right)+\mu\Phi \quad (5)$$

where $u, v, w$ are the fluid velocity components, $\rho$ is density, P is pressure, $\mu$ is the fluid dynamic viscosity, Cp is the specific heat capacity, T is temperature, and $\Phi$ is the dissipation function, which gives the time rate at which energy is dissipated per unit volume due to viscous effects. The density of the Krypton is calculated from the ideal gas law below, where $M_w$ is the molar mass of the Krypton and $R$ is ideal gas constant.

$$\rho=\frac{P}{(R/M_w)T} \quad (6)$$

To reduce the computational cost, only one quadrant of the sub-unit is modeled. At the symmetry surfaces, the gradient of all variables are set to zero. The numerical simulation was conducted using ANSYS Icepak, a commercially available CFD code based on the finite volume method. The three-dimensional governing equations are discretized by applying a finite volume method in which conservation laws are applied over finite-sized control volumes around grid points, and the governing equations are then integrated over the volume. Quick scheme was used to discretize convection/diffusion terms in momentum and energy equations. The numerical simulation is accomplished by using the SIMPLE algorithm. In this technique, using a guessed pressure field, the velocity components in three directions are first calculated from the Navier–Stokes equations, and then to satisfy the continuity equation, the pressure and velocities are corrected. The numerical solution is regarded as convergent at an iteration in which the summation of absolute values of relative errors of temperature, velocity components and pressure reach $10^{-9}$, $10^{-6}$ and $10^{-4}$, respectively. Convergence with respect to the number of grids in the computational domain is carefully checked and a fine grid is used in the regions of the boundary layers where the gradients of velocity and temperatures are steeper.

Two main factors influence the heat loss from the emitter: (i) the surface area of the emitter and W fins in contact with the gas, and (ii) the flow pattern and thermal boundary layer thickness near the hot emitter surfaces. The latter can enhance or diminish the heat loss if the thickness of thermal boundary layer decrease or increase because of flow circulation in the system due to the natural convection, respectively. Intuitively, larger emitter surface area and closer distance between MPV cells and the emitter result in more heat loss but it is not clear a *priori* how the size of W fins and MPV cell changes the flow pattern in the optical cavity and consequently the convective heat transfer near the hot surfaces. The effect of normalized fin length ($L_{fin}/D$) and MPV cells dimension ($L_{cell}/D$) on overall heat loss from the emitter is shown in Extended Data Figure 3B. The dimensions are normalized with respect to pipe diameter (D). Counterintuitively increasing the length of W fins decreases the heat loss from the emitter. This is because the convective enhancement of heat transfer near the surfaces reduces, while the change in the conductive heat contribution is minimal. As can be seen from temperature distribution in a plane 20 cm from the base of the system (Extended Data Figure 3A), for larger $L_{fin}/D$, the high temperature and low velocity hot gas is trapped between two adjacent fins causing a thicker thermal boundary layer near the hot surfaces to develop and consequently lower heat loss compared the case with small $L_{fin}/D$. The effect of $L_{cell}/D$ on heat loss is also shown in Extended Data Figure 3A. As seen by decreasing the MPV cell size, the heat loss decreases which is mainly due to the diminishing of conductive contribution of heat loss at larger cell-emitter distances. Clearly, for small distances between the emitter and cell the heat loss is dominated by the conduction and fluid circulation due to natural convection paly insignificant role. This is clearly shown in temperature contours in Extended Data Figure 3A as the thickness of thermal boundary layer doesn't change significantly by changing the size of MPV cells. In conclusion, the analysis suggests, (i) the convective heat loss from the system is <5 kW/m$^2$ of cell area (ii) increasing the emitter to cell area ratio not only enhances the radiative power incident onto the MPV cells but it also helps to minimize the convective heat losses by trapping hot gas between fins creating stagnant zones.

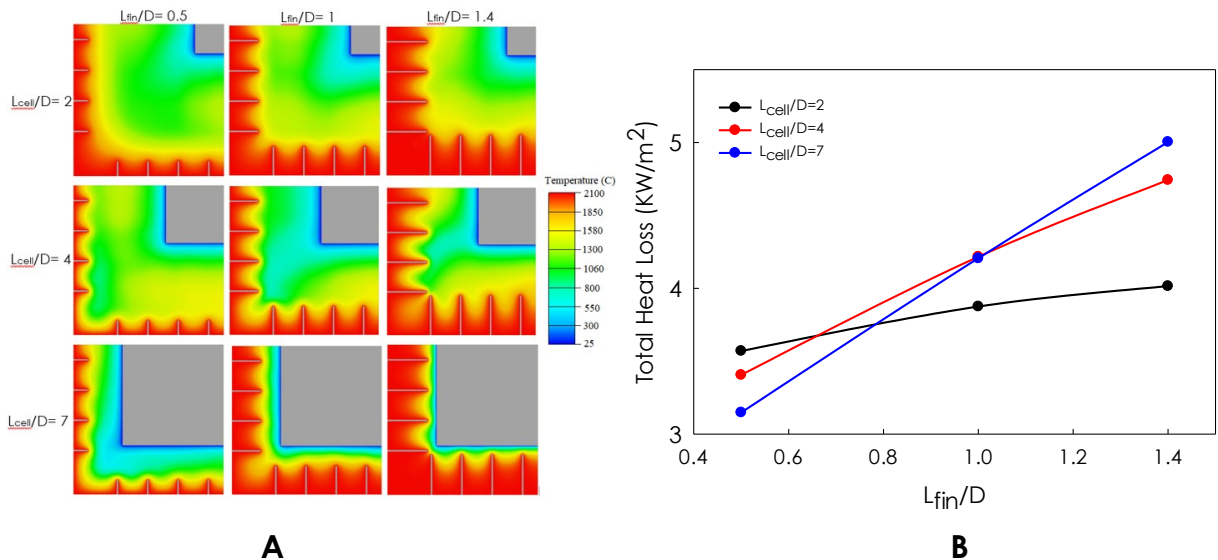

**Extended Data Figure 3: A –** Temperature distribution in a plane 20 cm from the base of the system **B –** effect of size of W fins and MPV cell on heat loss

# Techno-Economic Modeling

The full-scale system size is selected to approach the asymptotic minimum cost that would be achieved at infinite size, while remaining at a scale that could be reasonably manufactured (1 GWh-e), as shown in Extended Data Figure 4A. The system consists of two tanks of the same volume and wall thickness, with only one filled with Si, and the colder tank requires less insulation. The system is designed to have 10 hours of storage, with equal charging (resistive heating) and discharging (MPV power cycle) rates of 100 MW-e. A full-scale model of this system is presented in Extended Data Figure 4B, with tanks of ~15 m diameter.

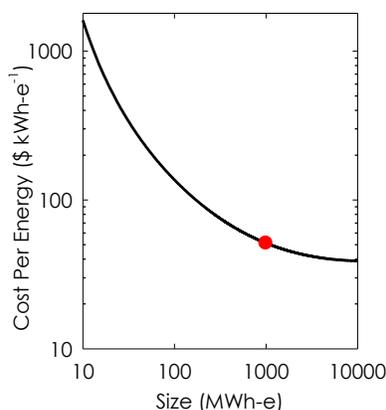
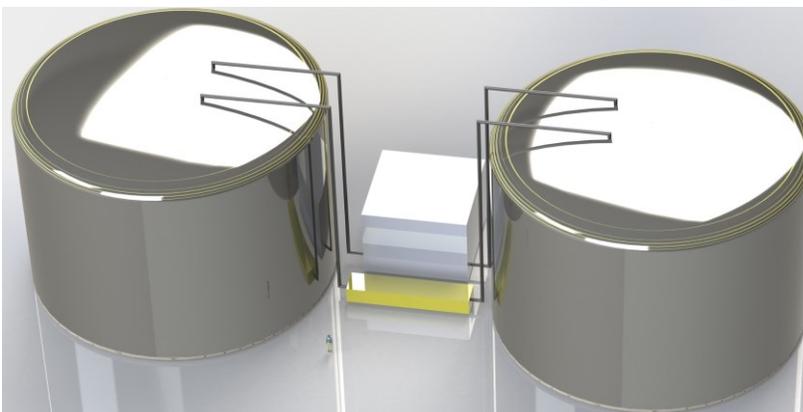

**Extended Data Figure 4A:** Scaling tradeoff between cost and size for TEGS CPE components. **B:** TEGS-MPV system concept consisting of a hot and cold tank, MPV heat engine (white), and resistive heating charger (yellow). System shown stores 1 GWh-e, with human shown for scale.

## Tanks

The material in direct contact with Si, providing mechanical and chemical containment, is made from isostatic graphite at a typical quoted cost for large quantities of $7/kg. This graphite grade has multiple trade names, such as KYM-30, AS-TJ, AR-06, and G330. The common features are a density greater than 1,750kg/m$^3$, with particle and pore size below 50 µm. These large tanks can be built in sections as shown in Extended Data Figure 5A, approximately one meter in size. The tank has two layers, which reduce the likelihood of leaks. The units are connected by flanges on all edges with high strength (120 MPa tensile) carbon fiber composite (CFC) threaded rod and nuts, as shown in Extended Data Figure 6B. As these CFC fasteners are exposed to Si, an experiment was conducted at 1800°C for 120 minutes to determine their behavior in contact with Si. As shown in Extended Data Figure 6C, similar to the bulk graphite tank, the threaded rod developed a SiC protective barrier, and retained its mechanical integrity. Sealing is achieved using a thin grafoil gasket described in the following proof of concept experiments.

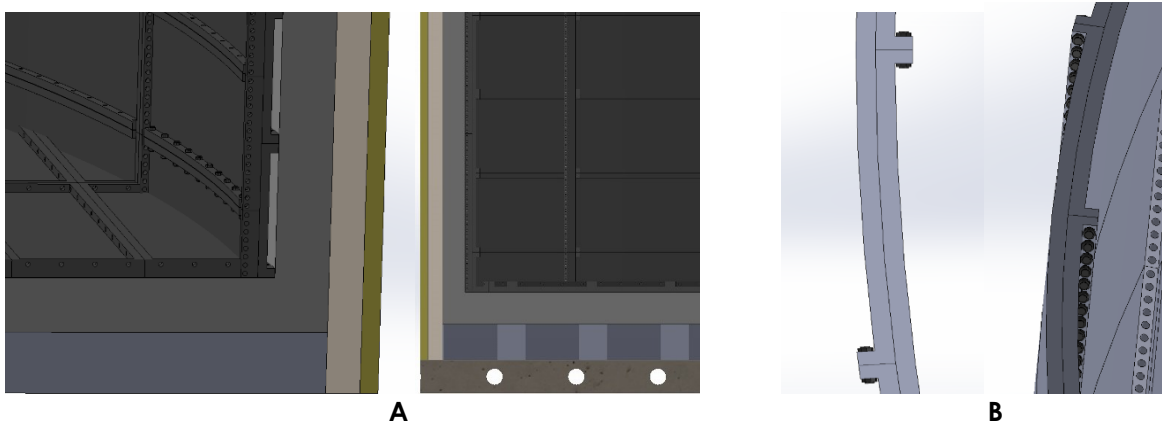

**Extended Data Figure 5A** – TEGS-MPV tank sections showing section and insulation design. **B** – Graphite tank sections and bolting scheme using two layers.

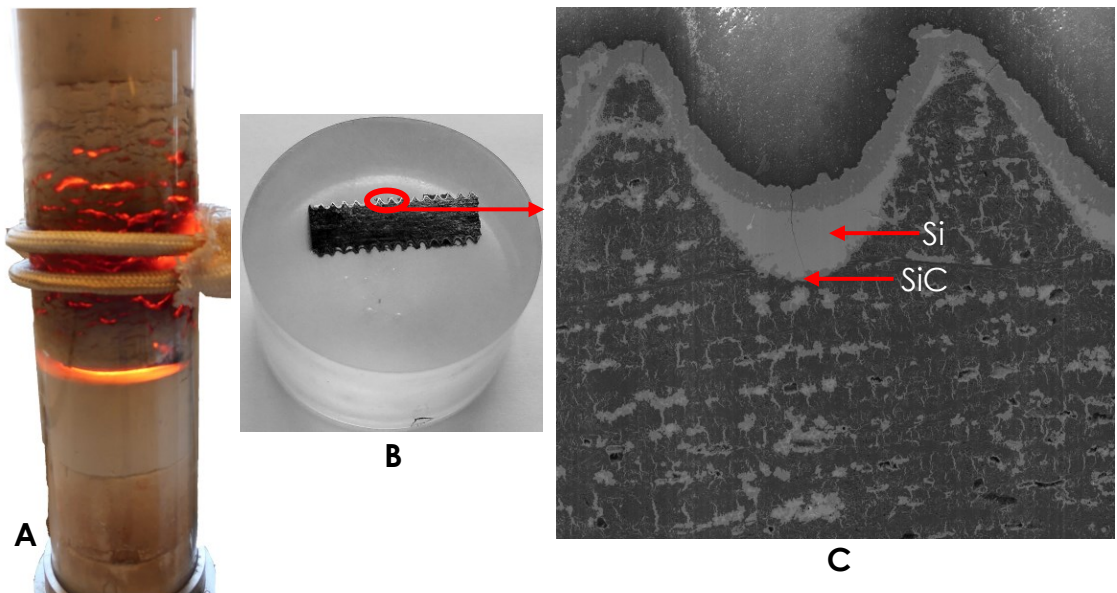

**Extended Data Figure 6A** – TEGS-MPV proof of concept scale experiment setup. **B** – Cross section of threaded rod after exposure to molten Si. **C** – SEM image of two teeth from the threaded rod showing SiC scale.

Mechanically, the tank wall thickness is designed to resist hoop stress with a minimum factor of safety of two, based on the tensile strength of graphite at room temperature (50 MPa), as shown in Equation 4. This safety factor increases with temperature, as the graphite strength increases with temperature[32]. The hoop stress in the wall decreases linearly with height, as the internal pressure arises from the gravitational force acting on the fluid, and this fact can reduce the graphite required by up to a factor of two.

$$t_c = \frac{(\rho g H) r}{(\sigma_{fracture} / SF)} \qquad (4)$$

$$t_{ins} = r_i \times \exp\left(\frac{2\pi H k (T_i - T_o)}{Q}\right) \tag{5}$$

The graphite tank is insulated with multiple layers of insulation, with the insulation thickness calculated using Equation 5 and the effective thermal conductivities listed in Extended Data Table 1. Immediately outside the graphite tank is graphite felt insulation, at a cost of $7,000/m³ based on multiple quotes. This material is used until the radial temperature drops to 1,350°C. At this point, a lower cost aluminum silicate ceramic fiber blanket is used at a cost of $400/m³. The cost of this material is low in part because it is widely used in ovens. Even lower in cost and thermal conductivity is fiberglass insulation, at $85/m³, so this material is used as an outer insulation layer below 540°C. The insulation cost is dominated by the graphite felt layer, which bridges a 1,000°C temperature decrease in the hot tank, but only 550°C in the cold tank. For this reason, the cold tank cost is $10/kWh-e cheaper than the hot tank.

The graphite tanks rest directly on rigid graphite insulation board, at a cost of $13,000/m³. Below 1,700°C, this board rests on a calcium aluminate based castable cement (WAM ALII HD), at a cost of $6,000/m³. This material is used for its compressive strength in a cinder block geometry, and the cavities are filled with aluminum silicate insulation to minimize cost, radiative heat loss, and natural convection. The castable cement then rests on a concrete foundation that can be cooled by forced air or water, as is the case in current molten salt CSP plants[43]. The concrete cost is $200/m³ and cooling cost is estimated $60/m² based on a designed heat flux of 400 W/m² and cost of recirculated cooling[44] of $80/kW. These tanks reside inside an inert atmosphere, achieved with a cold steel[45] shell.

Construction costs are estimated based on the cost of constructing molten salt CSP tanks[17], adding the cost to assemble additional components. For example, the cost to layout and bolt together the graphite tanks is estimated assuming that each section takes five minutes to position and one minute to install each bolt. With an estimated worker salary of $50,000 per year, the total tank construction cost comes out to $3.14/kWh-e.

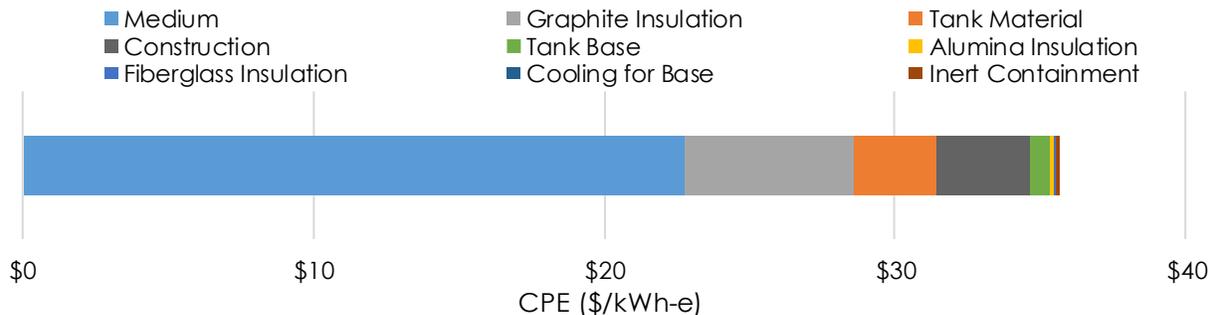

**Extended Data Figure 7** – Tank cost breakdown for 100% Si medium base case

### Heater

The heater consists of a 37 x 15 array of graphite pipes of 10 mm inner diameter and 20 mm outer diameter. Between columns of these pipes are graphite rods that are used as electric resistance heating elements as shown in Extended Data Figure 8. The heater is designed so that the peak heater temperature is 2500°C. The pricing for pipes and rods is from graphitestore.com,

at $100/m for pipes and $18/m for rods. A quote was obtained for these custom headers, at a cost of $0.25/kW. Thyristor based power supplies for the heaters were quoted at $5/kW. These supplies use silicon controlled rectifiers (SCR) to modulate power by rapidly switching, with efficiencies as high as 99.5%[33]. In a large heater such as this one, heating elements can be arranged in a series-parallel configuration to match the overall heater resistance to supplied voltage, thus minimizing or eliminating the need for voltage transformers. The heater is insulated using the same approach as the tanks, by restraining the heat flux loss to be the same 400 W/m$^2$ as the tanks. The total heater cost is only $0.02/W-e because it is so power dense and is driven by the power supply cost, pipe cost, and heating element cost. A breakdown of the heater cost is given in Extended Data Figure 9.

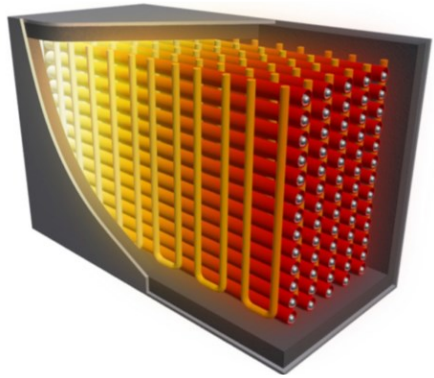

**Extended Data Figure 8:** Heater cross section showing horizontal graphite pipes connected to headers to transport Si and vertical graphite rods for resistive heating.

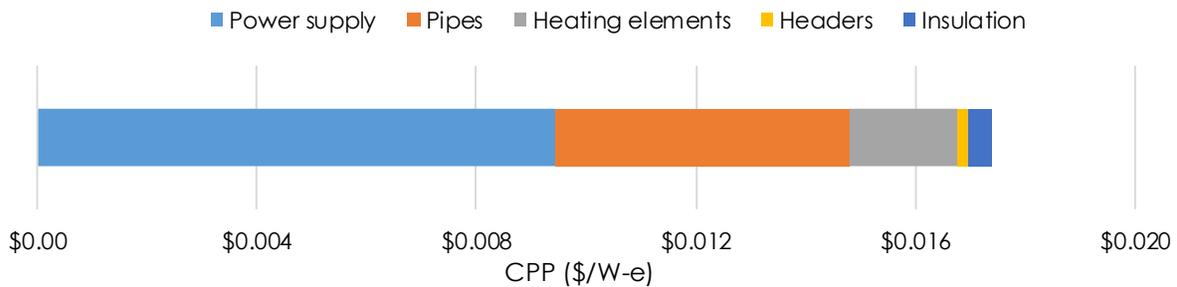

**Extended Data Figure 9** – Heater cost breakdown for the base case

## MPV Cost

The major variable in the power block cost is the MPV cell cost. It is expected that this cost will fall between the price of silicon PV cells[1] at $50/m$^2$ and the current cost of manufacturing GaAs cells at $10,000/m$^2$. The power density of this high temperature system is modeled assumed to be 100 kW/m$^2$, resulting in a cell cost between $0.001/W-e and $0.10/W-e. Nonetheless, here we have taken the more conservative upper limit on cost and lower limit on power density in our primary cost model. Another important cost is that of the inverters to convert the DC power to AC. These are priced at $0.08/W-e based on national averages[1] for central inverters in utility scale PV. Cooling of the MPV cells is priced at $0.08/W based[44] on recirculating cooling 8°C above ambient, where the required watts of cooling per watt of electricity generated is calculated as (1-RTE)/RTE. W foil, which is used as a vapor pressure barrier to suppress the evaporation of graphite was quoted at $700/m$^2$, or $0.035/W-e. The graphite piping and insulation is similar to the previously described systems and has a small effect on cost. The cost of constructing the MPV power block and heater are estimated by including other CPP construction costs based on previous analysis and adding the estimated labor cost to assemble additional components. For example, the time to install each pipe in the heater and MPV systems is estimated to be 10 minutes, and 30 minutes is estimated to install each unit of tungsten foil. Based on the labor rates discussed above, the construction cost of components that scale with power is estimated at $0.03/W-e. A breakdown of the MPV costs is given in Extended Data Figure 11. An overview and detail view of the MPV layout is shown in Extended Data Figure 10.

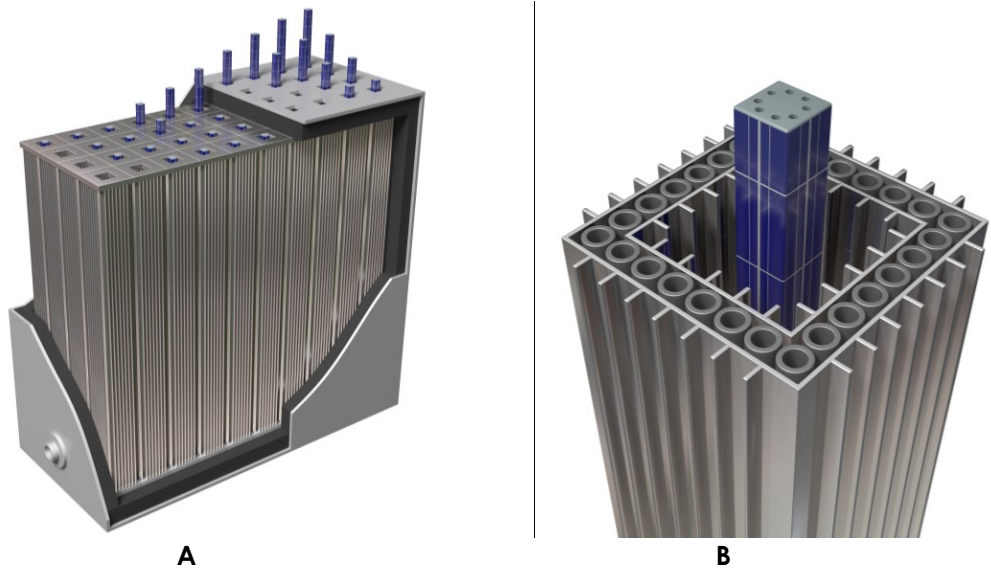

**Extended Data Figure 10A** – TEGS-MPV power block design. Vertically actuated MPV arrays are exposed to tungsten foil coated graphite pipe emitters. **B** – Single MPV unit, fins are used to increase effective emissivity of via the blackbody effect.

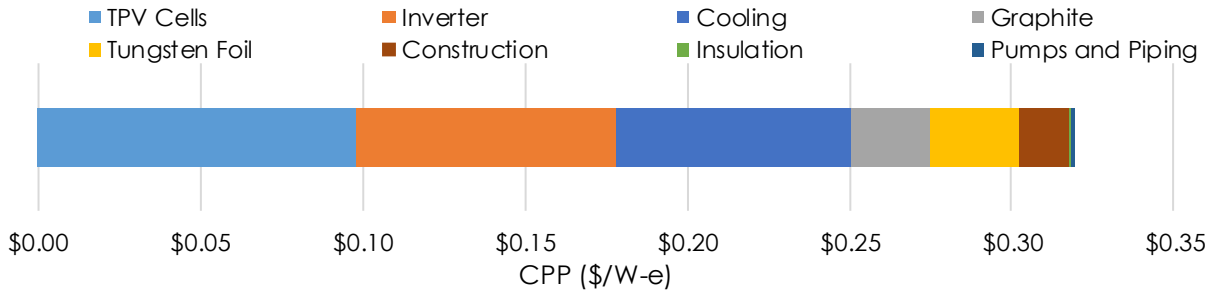

**Extended Data Figure 11** – MPV cost breakdown for the base case

## Pumps and piping

Because of the energy density of silicon, to discharge at 100MW-e, a flow rate of only 0.2 m$^3$/s (~3200 gpm) is needed. For perspective, this is similar in size to the water pumps found on fire engines. The pressure required is mostly to make up for gravitational head and will vary between 0.1-0.4 MPa (1-4 atm), which can easily be met with a centrifugal pump consuming ~ 40 kW of power. This flow rate and pressure can be met with a 330 mm (~ 1 ft) diameter centrifugal pump (SAE size 8x10-13), and the power requirement is negligible compared to the 100 MW-e power output, which a key advantage of using liquid Si/metal. The pump material is graphite, and the cost will be dominated by the 15 m shaft required to locate the pump in the bottom of the tank with the motor above the tank. The total mass of a pump is estimated at 2,000 kg and a pump is needed for each of the two tanks, and the graphite grade and cost match that of the tank material. The primary piping network between the tanks, heater, and MPV have a nominal diameter of 250 mm to minimize dynamic head loss. The cost of pumps and piping are included with components that scale with the power output (CPP) of the system and are shown in Extended Data Figure 11.

Extended Data Table 1: Material Costs and Properties

| Material | Density (kg/m3) | Cost ($/kg) | Thermal Conductivity (W/m-K) | Temperature Limit (°C) | Source |
|---|---|---|---|---|---|
| 553 Silicon | 2400 | 1.6 | 25 | 3250 | Quotes[46] |
| Isostatic Graphite | 1850 | 7.0 | 30 | 3600 | Quotes |
| Rigid Graphite Insulation | 24 | 540 | 0.3 | 2800 | Quotes[47] |
| Graphite Felt | 14 | 500 | 0.3 | 2800 | Quotes[47] |
| Aluminum Silicate | 100 | 4.0 | 0.2 | 1350 | Quotes[48] |
| Fiberglass Blanket | 12 | 7.1 | 0.05 | 540 | [49] |
| WAM ALII | 2700 | 2.2 | 1.5 | 1700 | Quotes[50] |
| Scrap steel | 7000 | 0.1 | 30 | 2862 | [51] |
| Tungsten Foil (0.1mm thick) | 19000 | 350 | 100 | 3400 | Quotes |

Extended Data Table 2: Effect of IRR on Max CPP

| | CPE | RTE | CPP | Life | Max CPP | | | |
|---|---|---|---|---|---|---|---|---|
| | | | | | 0% | 4% | 15% | 20% |
| TEGS-MPV | $36 | 50% | $0.34 | 30 | $3.06 | $1.62 | $0.37 | $0.18 |
| PHS | $60 | 90% | $0.75 | 30 | $4.90 | $2.60 | $0.61 | $0.31 |
| CAES | $27 | 75% | $0.60 | 30 | $4.46 | $2.49 | $0.77 | $0.52 |
| Li-ion | $150 | 90% | $0.08 | 10 | $0.33 | $0.01 | -$0.55 | -$0.71 |
| Lead-acid | $300 | 80% | $0.45 | 10 | -$1.34 | -$1.63 | -$2.14 | -$2.28 |
| Flywheel | $2900 | 60% | $0.30 | 30 | -$25.03 | -$26.69 | -$28.13 | -$28.34 |

## Alternative embodiments: Using solid storage

The usage of a liquid storage medium requires pumping, which could potentially be avoided if a solid storage medium were to be used. It is in this sense that very inexpensive forms of carbon exist that would have very low CPE values. However, in the preceding analysis, we have focused specifically on liquids/metals, because of the heat transfer issues that would arise from attempting to use a solid storage medium. It should first be noted that in general a gaseous storage medium will not have sufficiently high energy density to offer a competitive embodiment, since the atom or mass density of gases is generally 2-3 orders of magnitude lower than that of a solid or liquid. This generalization assumes the gas is nominally at 1 atm pressure, such that any vessel used to contain it need not become a pressure vessel, which would become extremely thick walled and cost prohibitive at the large grid scales of interest for solving the storage problem. This generalization also assumes the gaseous medium's energy content is based on its sensible heat and not a chemical reaction enthalpy. Assuming both of these are true, the generalization that a gaseous storage medium will not be competitive economically applies. This then leads one to consider a solid medium. Assuming the solid medium is not fluidized, in which case its density would decrease back to a value closer to a gas, the value of using a solid medium in a TEGS configuration would derive from the advantage of not having to move it. In this sense the idea is then to have a solid medium that consists of large blocks or alternative shapes and a key parameter becomes the surface area to volume ratio for the units of solid. Using graphite as an example solid, assuming a 500°C temperature swing (i.e., 1900-2400°C) with a nominal heat capacity of 720 J kg$^{-1}$ K$^{-1}$, density of 2270 kg/m$^3$ and a high temperature thermal conductivity of ~ 30 W m$^{-1}$ K$^{-1}$, a 100 MW plant with 10 hrs of storage would require storage of 10$^7$ kg of graphite. This could be stored in a roughly 18 m diameter 18 m tall cylindrical tank and if the MPV power density was ~100 kW/m$^2$ as has been required herein to reach the high efficiencies that enable the concept to compete with PH, then ~ 1000 m$^2$ of surface area would be required. This would then conceivably only require a small number (i.e., order 10) divisions of the graphite mass to make slots where the MPV cells could be inserted. This is important, because it would then establish the characteristic length over which the heat would need to be conducted during the discharge. Generally, for storage times of 10 hrs or greater this characteristic length is likely to be greater than 1 ft (~ 0.3 m). This then identifies a key problem with such an approach, which is that inherently, the heat must be conducted from the body of the volume to the surface where the radiation occurs. This is unavoidable in a situation where sensible heat is being used and the medium is stationary. Nonetheless, the order 100 kW/m$^2$ fluxes required to make the system efficient and cost effective will induce very significant thermal gradients and also thermal transients in the solid mass, especially near the surfaces.

For example, consider a solid block of graphite with an initially uniform temperature of 2400°C. If it is exposed to MPV cells that draw away a 100 kW/m$^2$ heat flux the surface will immediately cool as heat is conducted from the hotter portion of the block to the surface. In this case it is useful to estimate the temperature gradients that would develop, as well as the thermal transients as every material has intrinsic limits beyond which it will mechanically fracture and break into multiple parts. Conceptually, a mechanical fracture in such a system could prove catastrophic as a failure mechanism because it could cause a portion of the solid to fall or slide by gravity and directly contact an MPV cell, which would cause overheating. Nonetheless, the likelihood of such a failure can be assessed by approximating the solid block temperature profile as similar to that of a 1-D semi-infinite medium, for which a closed form solution exists[52] when there is a constant heat flux at the surface,

$$T(x,t) = T_i + \left(\frac{2q\sqrt{\alpha t/\pi}}{k}\right) \exp\left(\frac{-x^2}{4\alpha t}\right) - \left(\frac{qx}{k}\right) erfc\left(\frac{x}{2\sqrt{\alpha t}}\right) \qquad (6)$$

Where $T_i$ is the initial temperature, $q$ is the 100 kW/m$^2$ constant heat flux, $\alpha$ is the thermal diffusivity, $k$ is the thermal conductivity, $t$ is time and $x$ is the depth being evaluated (i.e., the distance to the surface). Using the properties for graphite mentioned above, this simple calculation reveals that, for example at a location 1 cm deep from the surface, the temperature will decrease at a rate greater than 40°C/sec throughout the first two minutes of discharge. Such an extreme transient would surely result in cracking and mechanical failure within 100 cycles (based on our own previous experiments [unpublished] thermal cycling and cracking graphite with 100°C/min heating rates). Furthermore, a large thermal gradient of ~ 3300K/m would need to exist at the surface. Such a high gradient is not problematic for a thin walled pipe, as would also occur in the liquid based TEGS-MPV embodiment discussed herein. This is because, for example, a 5 mm thick pipe wall would only experience a 15-20°C temperature drop across its wall. However, a solid storage medium would have to have a low surface area to volume ratio to keep the MPV cost from dramatically increasing, and for a characteristic length of ~ 1 ft, the temperature difference would need to be ~ 1000°C. Such a large temperature difference is problematic from an efficiency standpoint because the surface would have to be ~1000°C colder than the center of the storage medium during the majority of the discharge. This would then lead to a much lower efficiency that what has been predicted herein for emitter surface temperatures in the range of 1900-2400°C. It should be further noted that this issue of the transient buildup of conductive thermal resistance during discharge is rather fundamental and immutable. It is an intrinsic characteristic of relying on transient heat conduction through a solid for the discharge. Although the cost of solid storage media could potentially be very low, the thermal management issues that would arise during the discharge are daunting, and likely insurmountable. It is for this reason that a liquid storage medium is highly preferred and likely the only way to realize a system with sufficiently high efficiency that it can enable eventual 100% penetration of renewables.